**Geometry-Controlled Exciton Selectivity in Monolayer $MoS_2$ Using Plasmonic Hollow Nanocavities**

Abdullah Efe Yildiz[1,2] and Emre Ozan Polat[1,2,*]

[1]*Department of Physics, Bilkent University, 06800, Ankara, Turkey*

[2]*UNAM - National Nanotechnology Research Center and Institute of Materials Science and Nanotechnology, Bilkent University, Ankara 06800, Turkey*

**Corresponding author: emre.polat@bilkent.edu.tr*

## Abstract

Spectral control of closely spaced excitonic transitions is central to valleytronic photonics, nanoscale light sources, and wavelength-encoded sensing. In monolayer molybdenum disulfide ($MoS_2$), the A and B excitons are separated by only tens of meV, making selective excitonic emission control both fundamentally important and technologically challenging. Here, we numerically investigate plasmon-enhanced excitonic emission from monolayer $MoS_2$ coupled to vertically oriented hollow gold nanocylindrical cavities through a dielectric spacer. Finite-difference time-domain simulations combined with a photoluminescence-rate framework enable separate evaluation of excitation enhancement, radiative decay modification, nonradiative quenching, and excitonic charge generation. By tuning the cavity aspect ratio, the localized surface plasmon resonance is selectively aligned with either the A- or B-exciton transition, while the spacer thickness and refractive index regulate near-field coupling and the local density of optical states. Under optimized conditions, the excitation rate is enhanced by up to 4.34-fold and the radiative decay rate by more than 40-fold, yielding photoluminescence enhancements of 143.85 and 87.27 for the A and B excitons, respectively. The cavity also redistributes the relative excitonic peak intensities, producing exciton-selective peak ratios up to 2.4 times higher than those of bare $MoS_2$. These results establish hollow plasmonic nanocavities as geometry-tunable platforms for exciton-selective emission and charge-generation control in atomically thin semiconductors.

## Introduction

Two-dimensional transition metal dichalcogenides (TMDCs) provide a unique platform for nanoscale optoelectronics owing to their strong light-matter interactions and tightly bound excitons that remain stable at room temperature[1,2] . In monolayer (ML) $MoS_2$, the direct bandgap at the K and K′ valleys together with spin-orbit splitting of the valence band gives rise to spectrally distinct A and B excitons[3-5]. These excitonic transitions dominate the optical response and underpin applications in light-emitting devices, photodetectors, and valleytronic systems[6-8] . However, despite their large oscillator strengths, the atomic thickness of ML $MoS_2$ fundamentally limits optical absorption and emission[9,10]. Rapid nonradiative decay and weak out-coupling to the far field suppress the external photoluminescence (PL) quantum yield, creating a practical bottleneck for device performance[11,12]. Engineering the local electromagnetic environment is therefore essential to translate intrinsic excitonic strength into experimentally measurable emission and photocurrent enhancement[13,14].

Plasmonic nanostructures offer a scalable and experimentally accessible route to manipulate emission by concentrating electromagnetic fields and modifying the local density of optical states (LDOS)[15,16.] Gold nanoantennas and gap resonators have demonstrated enhanced PL and excitation-rate amplification via near-field confinement and Purcell-enhanced radiative recombination[17,18]. Yet, the literature generally focuses on planar or solid metallic elements integrated with $MoS_2$, that predominantly generate lateral hotspots, limiting the control over out-of-plane confinement and three-dimensional cavity modes[14-20]. Hollow and coaxial metallic nanocavities, readily fabricated via template-assisted growth or galvanic replacement, support hybrid radial-longitudinal plasmon modes with deeply subwavelength mode volumes and geometry-dependent spectral tunability[21,22]. Despite their established synthesis and strong electromagnetic field confinement, hollow plasmonic nanocavities have only been sparsely explored for exciton selectivity in atomically thin semiconductors.

Here, we present a vertically oriented hollow gold nanocavity platform integrated with a dielectric spacer on ML $MoS_2$ on a device substrate ($Si/SiO_2/MoS_2$). Our structure is specifically tailored for geometry-controlled exciton-plasmon coupling. The coaxial cavity supports confined out-of-plane plasmon modes whose spectral position can be tuned through aspect ratio and wall thickness to overlap selectively with either the A or B exciton. Using finite-difference time-domain (FDTD) simulations combined with a PL and excitonic charge-generation analysis directly linked to experimentally measurable PL and carrier-generation rates, we systematically investigate the influence of cavity geometry and spacer thickness ($Al_2O_3$ or Polymethylmethacrylate (PMMA)). We demonstrate geometry-controlled enhancement of distinct excitonic transitions together with corresponding increases in electron-hole pair generation rates. While our numerical results provide an applicable pathway toward spectrally programmable emission and photocurrent control, introducing hollow plasmonic nanocavities as a geometry-addressable degree of freedom for exciton selection establishes a fundamentally new three-dimensional cavity paradigm for transition-specific control of light emission and charge generation in atomically thin semiconductors.

## Results

**Figure 1** demonstrates the geometric tunability and spectral selectivity of the gold hollow nanocylinder (AuHNC) platform designed for exciton-plasmon coupling with ML $MoS_2$. The investigated structure is shown in **Figure 1a**, consisting of a hollow AuHNC with height $H$, outer radius $R_0$, and inner radius $R_i$ placed on a dielectric spacer. To determine the spectral response of the cavities prior to integration with ML $MoS_2$, extinction cross-section spectra were calculated under broadband plane-wave excitation (400-800 nm). To better elucidate the spatial nature of the nanocavity modes, we analyzed vertical x-z cross-sectional electric-field enhancement maps extracted from the three-dimensional FDTD simulations. As shown in **Figures 1 b,c**, the localized surface plasmon resonance (LSPR) mode exhibits strong near-field localization around the hollow cavity and spacer region, with pronounced field variation along both the lateral and vertical directions of the cross-section. These maps reveal the out-of-plane confinement of the plasmonic mode and its sensitivity to the dielectric spacer. Upon introducing the spacer layer, the electric field becomes more strongly confined toward the inner cavity of the AuHNC, leading to an increased field intensity within the hollow region. This out-of-plane dependency of the mode forms the basis of the plasmon-exciton interaction yielding exciton selectivity within the A and B excitons in the weak-coupling regime.

**Figure 1d** shows the spectral tuning due to the increasing the effective cavity aspect ratio ($CAR$), which reflects the effective cavity depth relative to the wall thickness that governs hybrid radial-longitudinal plasmon confinement,

$$CAR = \frac{H}{R_0 - R_i} \tag{1}$$

$CAR$ serves as a control parameter for a pronounced redshift of the LSPR. This behavior arises from geometry-driven plasmon hybridization between radial and axial cavity modes as depicted in **Figures 1b, c**. The hollow configuration supports coupled charge oscillations on both the inner and outer metal surfaces, resulting in enhanced near-field confinement compared to solid gold nanostructures (**Supplementary Figure S1**). This dual interface mode coupling provides an additional degree of freedom for spectral engineering.

From this parametric study, two optimized geometries were selected to target the intrinsic excitonic resonances of ML $MoS_2$: $100 \times 50 \times 30\, nm\, (CAR = 5)$ aligned with the A exciton $100 \times 50 \times 20\, nm\, (CAR = 3.33)$ aligned with the B exciton. These dimensions ensure controlled spectral overlaps while preserving moderate linewidths suitable for weak-coupling conditions (**Figures 1e, f**). This design strategy can also be extended to other TMDCs whose excitonic transitions spectrally overlap with the LSPR of the AuHNCs. For example, AuHNCs with $CAR = 1.66$ can be used to target the B exciton of $MoSe_2$, while AuHNCs with $CAR = 4$ can be employed to enhance the A exciton of $WS_2$[23].

**Figures 1e** and **f** show the effect of spacer thickness on the plasmonic response of the optimized AuHNC geometries on $Al_2O_3$ and PMMA spacers. As the spacer thickness increases from 5 to 25 nm, the dominant LSPR mode exhibits a clear and monotonic redshift in both dielectric systems. This behavior reflects the progressive reduction of plasmon-substrate coupling and the effective modification of the local dielectric environment experienced by the near field.

For the A-exciton-targeted AuHNC geometry ($100 \times 50 \times 30\, nm\, (CAR = 5)$), the LSPR peak exhibits a pronounced thickness-dependent redshift. Across the investigated spacer thickness range (5-25 nm), the resonance shifts by 54.4 nm for $Al_2O_3$ and 26.9 nm for PMMA spacers (**Figures 1 e, f** and **Supplementary Figure S1**). A comparable systematic redshift is also observed for the B-exciton-targeted geometry, where the LSPR peak shifts by 49.7 nm for $Al_2O_3$ and 19.3 nm for PMMA (see **Table 1**).

In both cases, the larger spectral shift observed for $Al_2O_3$ arises from its higher refractive index relative to PMMA, which enhances the dielectric contrast at the metal-dielectric interface and increases the near-field sensitivity of the plasmonic mode. This thickness-dependent spectral evolution is consistent with a simplified sensing-volume model, in which the plasmon resonance wavelength scales with the effective refractive index experienced within the evanescent near-field decay length[24,25]:

$$\Delta\lambda = m\Delta n\left[1 - e^{-2d/l_d}\right] \quad (2)$$

where $d$ is the spacer thickness, $l_d$ is the evanescent field decay length, $m$ is the bulk refractive-index sensitivity of the plasmon mode, and $\Delta n$ represents the refractive-index contrast introduced by the spacer. The exponential term captures the finite spatial extent of the plasmonic near field, leading to saturation behavior at larger separations. In contrast to classical particle-on-mirror nanocavities[26], where the spacer controls metal-metal gap-mode hybridization, the present dielectric-substrate-supported hollow nanocylinder exhibits resonance tuning through near-field sampling of the spacer/substrate optical environment as shown in **Equation 2**. The hollow cavity and lower rim geometry increase the overlap between the LSPR near field and the dielectric spacer so that varying the spacer thickness from 5 to 25 nm modifies the effective dielectric screening around the plasmon mode. On the other hand, the secondary high-energy resonance wavelength remains largely insensitive to spacer thickness, exhibiting only minor shifts of a few nanometers.

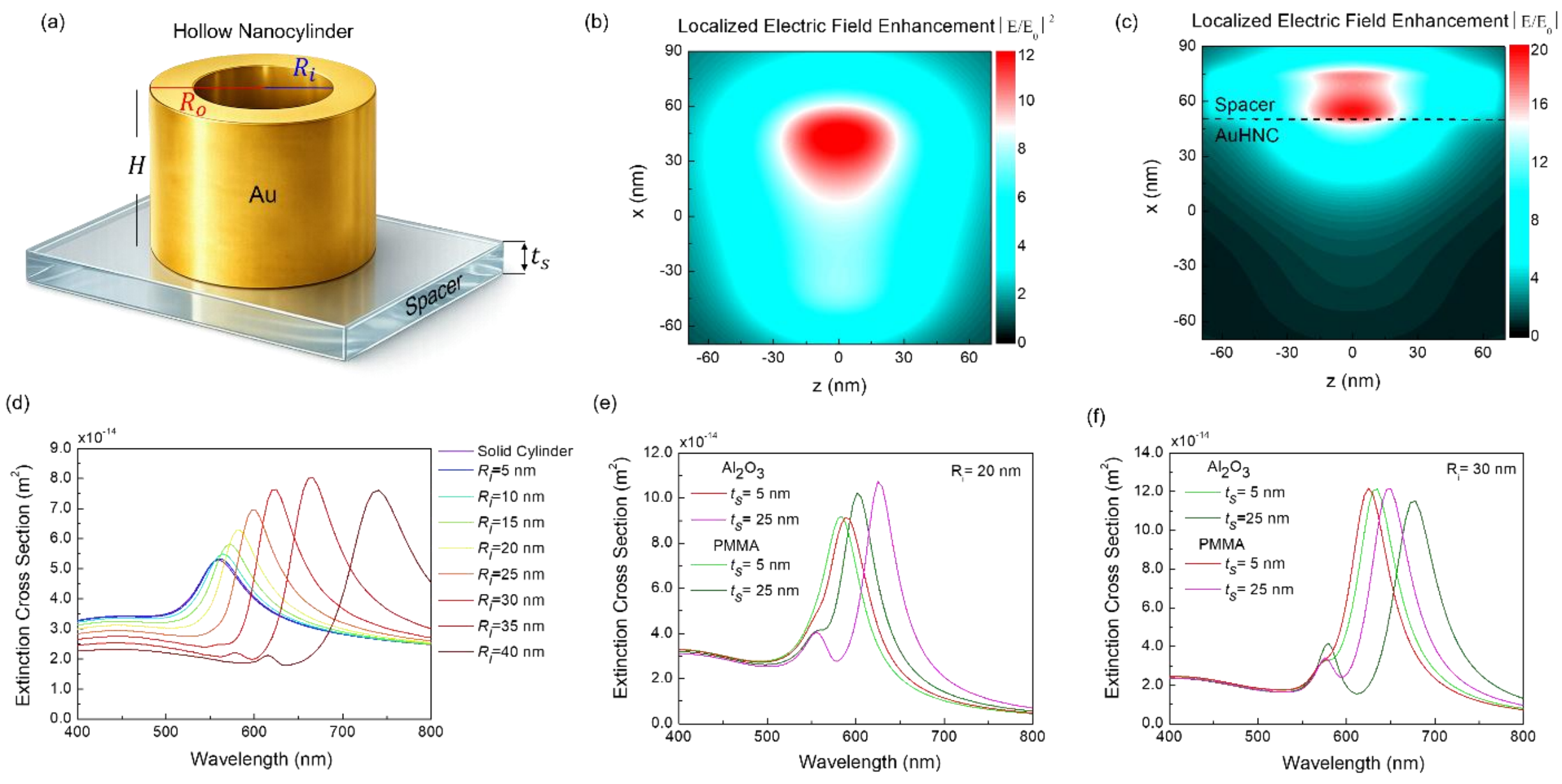

**Figure 1. Geometry-controlled plasmonic tuning and spacer-dependent LSPR evolution in gold hollow nanocylinders (AuHNCs). (a)** Schematic illustration of an AuHNC with height H, outer radius $R_0$, and inner radius $R_i$ positioned on a dielectric spacer. **(b,c)** Local electric-field enhancement maps for the AuHNC without spacer and with spacer, respectively, for a cavity aspect ratio of *CAR* = 3. The maps reveal strong field confinement near the cavity/spacer region and illustrate the out-of-plane localization of the plasmonic mode. **(d)** Calculated extinction cross-section spectra under broadband illumination from 400 to 800 nm for different AuHNC geometries. The systematic redshift of the dominant LSPR with increasing CAR enables geometry-based spectral tuning toward selected excitonic transitions. **(e,f)** Extinction cross-section spectra of the optimized AuHNC geometries with $R_i$ = 20 nm and $R_i$ = 30 nm, targeting the B- and A-exciton spectral regions of ML $MoS_2$, respectively. For both spacer materials, increasing spacer thickness systematically shifts the primary LSPR mode, whereas the higher-energy secondary mode shows only a weak dependence on spacer thickness. This indicates that the lower-energy mode is more strongly affected by spacer-mediated changes in the local dielectric environment and near-field distribution.

**Table 1. Spacer-thickness-dependent evolution of the LSPR peaks of AuHNC relative to the A- and B-excitonic transitions of ML $MoS_2$**. The table summarizes the LSPR peak wavelengths, the corresponding spectral shifts with respect to bare AuHNCs, and the full width at half maximum (FWHM) values for two spacer materials ($Al_2O_3$ and PMMA) with thicknesses ranging from 5 to 25 nm.

| $R_i = 30\ nm$ | LSPR Peak, $\lambda_1$ (nm) | | Peak Shift, $\Delta\lambda$ (nm) | | FWHM (nm) | |
|---|---|---|---|---|---|---|
| Spacer Thickness (nm) | $Al_2O_3$ | PMMA | $Al_2O_3$ | PMMA | $Al_2O_3$ | PMMA |
| 5 | 634.9 | 625 | 11.4 | 1.5 | 58.9 | 58.4 |
| 10 | 650.4 | 634.9 | 26.9 | 11.4 | 59.6 | 59 |
| 15 | 661.2 | 640 | 37.7 | 16.5 | 60.2 | 58.7 |
| 20 | 666.7 | 645.1 | 43.2 | 21.6 | 61.4 | 59 |
| 25 | 677.9 | 650.4 | 54.4 | 26.9 | 62.1 | 59.4 |
| $R_i = 20\ nm$ | LSPR Peak, $\lambda_1$ (nm) | | Peak Shift, $\Delta\lambda$ (nm) | | FWHM (nm) | |
| Spacer Thickness (nm) | $Al_2O_3$ | PMMA | $Al_2O_3$ | PMMA | $Al_2O_3$ | PMMA |
| 5 | 588.2 | 584 | 5.5 | 1.3 | 72.3 | 66.6 |
| 10 | 601.5 | 588.2 | 18.8 | 5.5 | 61.9 | 65.9 |
| 15 | 615.4 | 597 | 32.7 | 14.3 | 57 | 61.3 |
| 20 | 620.2 | 601.5 | 37.5 | 18.8 | 53.1 | 57.6 |
| 25 | 632.4 | 602 | 49.7 | 19.3 | 51.5 | 55.5 |

This weak dependence indicates that this mode is dominated by localized charge oscillations confined to the cavity walls, with limited interaction volume extending into the surrounding dielectric. Consistent with this interpretation, the absence of significant linewidth broadening confirms that the observed spectral shifts originate primarily from dielectric screening effects rather than modifications of plasmon damping or radiative losses.

Next, we integrated the AuHNC plasmonic cavity system with ML $MoS_2$. **Figure 2** presents the absorption spectra of the AuHNC- $MoS_2$ hybrid system for $Al_2O_3$ and PMMA spacers with varying thicknesses, revealing the combined contributions of plasmonic and excitonic absorption processes. The investigated AuHNC array has an in-plane periodicity of 200 nm, and the corresponding unit-cell configuration is shown in **Figure 2a**.

The absorption enhancement is primarily governed by strong near-field confinement associated with the primary LSPR mode. As shown in **Figure 2b**, the electric-field enhancement map evaluated at the $MoS_2$ plane reveals pronounced in-plane localization around the hollow nanocavity, confirming efficient spatial overlap between the plasmonic near field and the monolayer $MoS_2$. This local excitation-field enhancement substantially amplifies excitonic absorption despite the atomically thin nature of $MoS_2$. While **Figures 1b,c** demonstrated the vertical confinement of the LSPR mode through x-z cross-sectional field maps, **Figure 2b** shows the corresponding in-plane near-field distribution evaluated directly at the $MoS_2$ plane. These complementary views demonstrate that the nanocavity mode is vertically confined around the spacer region and laterally overlaps with the monolayer semiconductor, thereby enhancing the local excitation field responsible for excitonic absorption.

**Figure 2c** establishes the dielectric-dependent plasmon-exciton interaction by comparing the absorption spectra of bare ML $MoS_2$ with those of the hybrid structure comprising $MoS_2$, spacer, and AuHNC for $Al_2O_3$ and PMMA spacers. Here, the AuHNC cavity geometry is defined by an inner radius of $R_i = 20$ nm, corresponding to an aspect ratio optimized for spectral proximity to the B-exciton transition. In the absence of any nanocavity, ML $MoS_2$ exhibits weak but distinct A- and B-excitonic absorption features. Upon integration with the AuHNC and dielectric spacer, a pronounced absorption enhancement emerges near the B-exciton resonance. Importantly, the magnitude and spectral alignment of this enhancement differ for $Al_2O_3$ and PMMA, reflecting the influence of the spacer refractive index on the LSPR energy and the resulting plasmon-exciton overlap. The higher refractive index of $Al_2O_3$ induces a stronger redshift of the LSPR compared with PMMA, leading to improved spectral alignment with the B-exciton and consequently stronger absorption enhancement.

**Figures 2d** and **2e** further investigate the spacer-thickness-dependent absorption response of the hybrid structure for $R_i = 30$ nm, corresponding to the A-exciton-targeted AuHNC geometry, using $Al_2O_3$ and PMMA spacers, respectively. For thinner spacers, particularly at $t_s = 5$ nm, ML $MoS_2$ resides within the high-intensity evanescent near-field region of the AuHNC, yielding pronounced enhancement of excitonic absorption. As the spacer thickness increases to 25 nm, the enhancement progressively diminishes due to the decay of the plasmonic near field away from the nanocavity, while the intrinsic exciton energies remain spectrally fixed.

In both A- and B-exciton targeted geometries, the broadband absorption background originates primarily from Ohmic dissipation in gold, whereas the enhanced and spectrally localized features arise from plasmon-mediated field concentration at the $MoS_2$ plane. Spacer thickness further modulates this effect by regulating the spatial overlap between the evanescent plasmonic field and the excitonic layer, as also shown in **Supplementary Figure S2**. In addition, the different refractive indices of $Al_2O_3$ and PMMA modify the LSPR position and near-field distribution, resulting in dielectric-dependent absorption enhancement and spectral overlap with the excitonic transitions. This behavior confirms that the system operates in the weak-coupling regime, where selective excitation enhancement is governed by spectral alignment and LDOS modulation rather than plasmon-exciton hybridization, as evidenced by the absence of excitonic energy shifts or Rabi splitting in the spectra[27,28].

To directly quantify plasmon-enhanced excitation in the hybrid structures discussed in **Figure 2**, we evaluate the charge generation rate (*CGR*) in ML $MoS_2$, which corresponds to the rate of electron-hole pair generation induced by optical absorption. In the numerical framework, *CGR* is computed using the imaginary dielectric function of ML $MoS_2$, as detailed in the Methodology section. The *CGR* is extracted using an in-plane electric-field monitor positioned at the $MoS_2$ layer so that only the tangential electric-field components responsible for carrier generation are included.

**Figure 3** summarizes the exciton-resolved *CGR* response of the AuHNC- $MoS_2$ hybrid system. The periodic AuHNC array integrated with the spacer- $MoS_2$-substrate stack is illustrated in **Figure 3a**. Two nanocavity geometries are considered, which were previously shown in **Figure 2** to spectrally align the LSPR with the B-exciton and A-exciton transitions of $MoS_2$, respectively. For the $R_i$ = 20 nm cavity, shown in **Figure 3b**, the plasmonic resonance overlaps predominantly with the B-exciton transition, leading to a pronounced increase in the *CGR* at this wavelength compared with bare $MoS_2$. In contrast, the enhancement near the A-exciton remains comparatively weaker, indicating that this geometry preferentially amplifies B-exciton charge generation. The enhancement is strongest for the smallest spacer separations, where the $MoS_2$ layer resides within the intense evanescent plasmonic near field.

When the cavity inner radius is increased to $R_i$ = 30 nm, the plasmonic resonance redshifts and becomes spectrally aligned with the A-exciton transition, as shown in **Figure 3c**. As a result, the *CGR* spectrum exhibits a strong enhancement around the A-exciton wavelength, whereas the B-exciton enhancement becomes relatively weaker. The inset of **Figure 3c** compares the CGR spectra for $Al_2O_3$ and PMMA spacers at a fixed thickness of 2 nm, allowing the influence of the dielectric environment to be examined independently of spacer separation. The spectra show that both spacer materials produce nearly identical excitonic peak positions, confirming that the excitonic resonances are intrinsic to the $MoS_2$ layer. However, the $Al_2O_3$ spacer yields a slightly stronger *CGR* enhancement than PMMA. This behavior originates from the higher refractive index of $Al_2O_3$, which induces a larger redshift of the plasmonic resonance and improves its spectral overlap with the A-exciton transition. Consequently, the plasmon-assisted excitation enhancement becomes marginally stronger for the $Al_2O_3$ spacer under otherwise identical geometrical conditions. This spectral redistribution confirms that the plasmonic cavity geometry can be used to selectively enhance different excitonic transitions in ML $MoS_2$.

The exciton-specific enhancement factors extracted from the *CGR* spectra are summarized in **Figures 3d and 3e**, where the results for $Al_2O_3$ and PMMA spacers are shown separately.

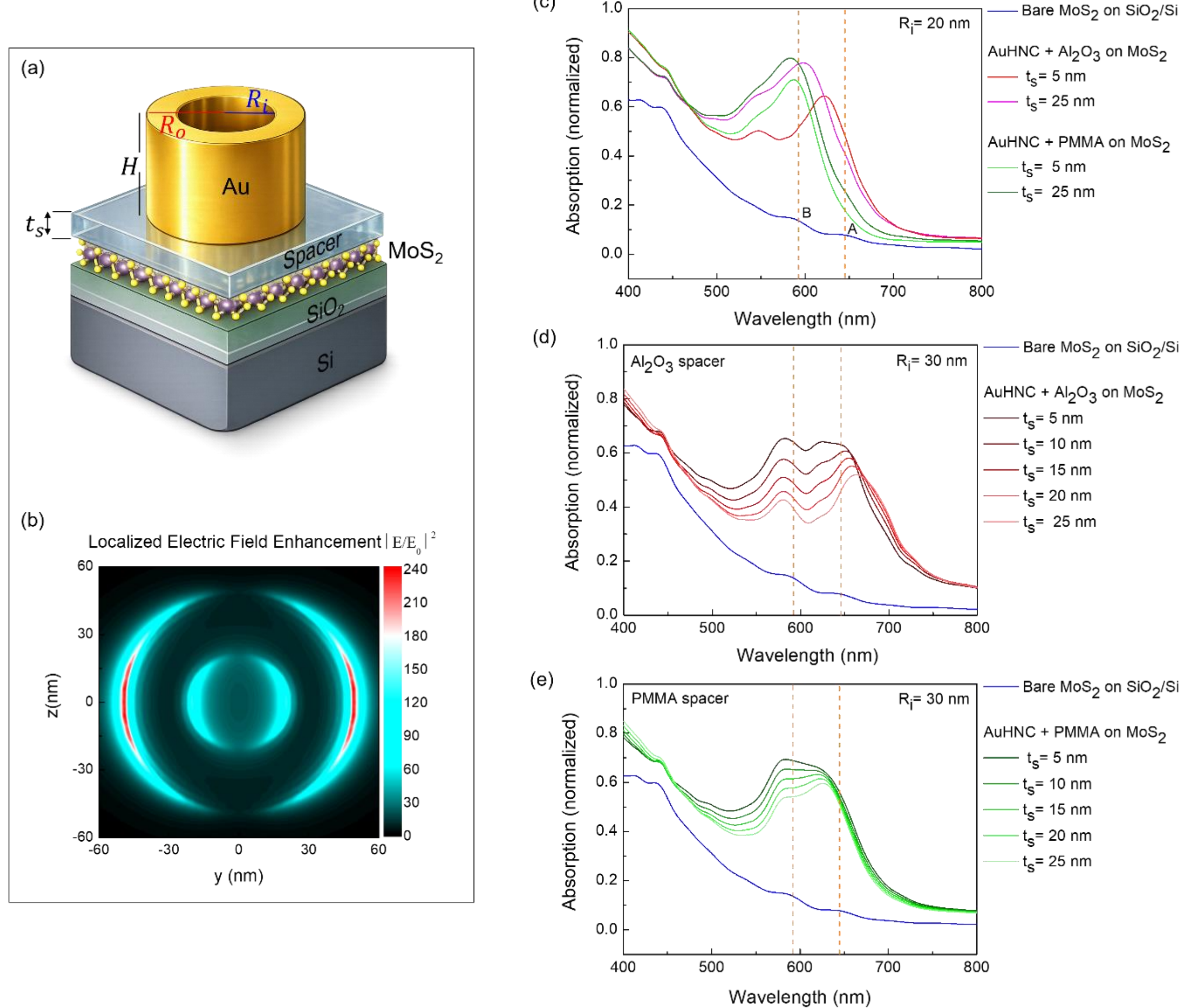

**Figure 2. Exciton-aligned absorption response of AuHNC–$MoS_2$ hybrid nanostructures.** **(a)** Schematic of the simulated unit cell consisting of ML $MoS_2$ on a $SiO_2$/Si substrate, separated from a periodic AuHNC array with 200 nm pitch by a dielectric spacer of thickness $t_s$. The AuHNC geometry has fixed outer dimensions and a variable inner radius $R_i$. **(b)** Local electric-field enhancement map evaluated at the $MoS_2$ plane, showing in-plane plasmonic near-field localization around the hollow AuHNC and its spatial overlap with ML $MoS_2$. **(c)** Absorption spectra of bare ML $MoS_2$ and AuHNC-spacer-$MoS_2$ hybrid structures with $R_i$ = 20 nm for $Al_2O_3$ and PMMA spacers, corresponding to the B-exciton-targeted geometry. **(d,e)** Spacer-thickness-dependent absorption spectra for the $R_i$ = 30 nm geometry, targeting the A-exciton region, with **(d)** $Al_2O_3$ and **(e)** PMMA spacers. Varying the spacer thickness controls the near-field overlap with ML $MoS_2$ and hence the excitonic absorption enhancement. The fixed exciton energies, together with spacer-dependent LSPR background and enhancement amplitude, indicate weak-coupling behavior and dielectric-dependent plasmon-exciton spectral overlap.

For the $R_i$ = 20 nm geometry, the B-exciton enhancement reaches 3.94 for $Al_2O_3$ and 3.67 for PMMA at the smallest spacer thickness of 2 nm, as shown in **Figure 3d**. As the spacer thickness increases, the enhancement gradually decreases for both spacer materials, approaching unity at larger separations as the plasmonic near-field interaction weakens. In contrast, for the $R_i$ = 30 nm cavity, the A-exciton enhancement reaches 4.34 for $Al_2O_3$ and 4.10 for PMMA at $t_s$ = 2 nm, as shown in **Figure 3e**. The enhancement again decreases with increasing spacer thickness due to the spatial decay of the plasmonic near field sampled at the $MoS_2$ plane. Taken together, these results demonstrate that the inner radius of the AuHNC cavity provides a direct geometrical handle for selectively enhancing different excitonic charge-generation channels in ML $MoS_2$.

Consequently, the plasmon-exciton coupling strength becomes marginally stronger for the $Al_2O_3$ spacer under otherwise identical geometrical conditions. As a result, the *CGR* spectrum exhibits a strong enhancement around the A-exciton wavelength while the B-exciton enhancement becomes relatively weaker (**Supplementary Figure S3**). This spectral redistribution confirms that the plasmonic cavity geometry can be used to selectively enhance different excitonic transitions in ML $MoS_2$.

Although the *CGR* enhancement quantifies the efficiency of plasmon-assisted exciton generation, it does not directly determine the observed PL intensity. The detected PL depends on both the excitation rate and the balance between radiative and non-radiative decay processes. In plasmonic environments, near-field enhancement can simultaneously increase excitation while also introducing additional non-radiative energy-transfer pathways to the metal. Therefore, achieving net PL enhancement (*PLE*) requires optimizing both the excitation enhancement and the emission efficiency.

For materials with intrinsically low quantum yield such as ML $MoS_2$ ($\eta_0 \approx 10^{-3}$-$10^{-4}$), the *PLE* can be approximated as[29,30]

$$PLE \approx \left|\frac{G}{G_0}\right| \times F_{rad} \tag{3}$$

where $G/G_0$ represents the *CGR* (excitation) enhancement factor and $F_{rad}$ is the normalized radiative decay rate of the hybrid system. The radiative and non-radiative decay rates are calculated using an in-plane dipole model for the $MoS_2$ exciton interacting with the plasmonic nanocavity, allowing the quantum efficiency of the hybrid structure to be expressed as:

$$QE = \frac{F_{rad}}{F_{rad}+F_{nonrad}} \tag{4}$$

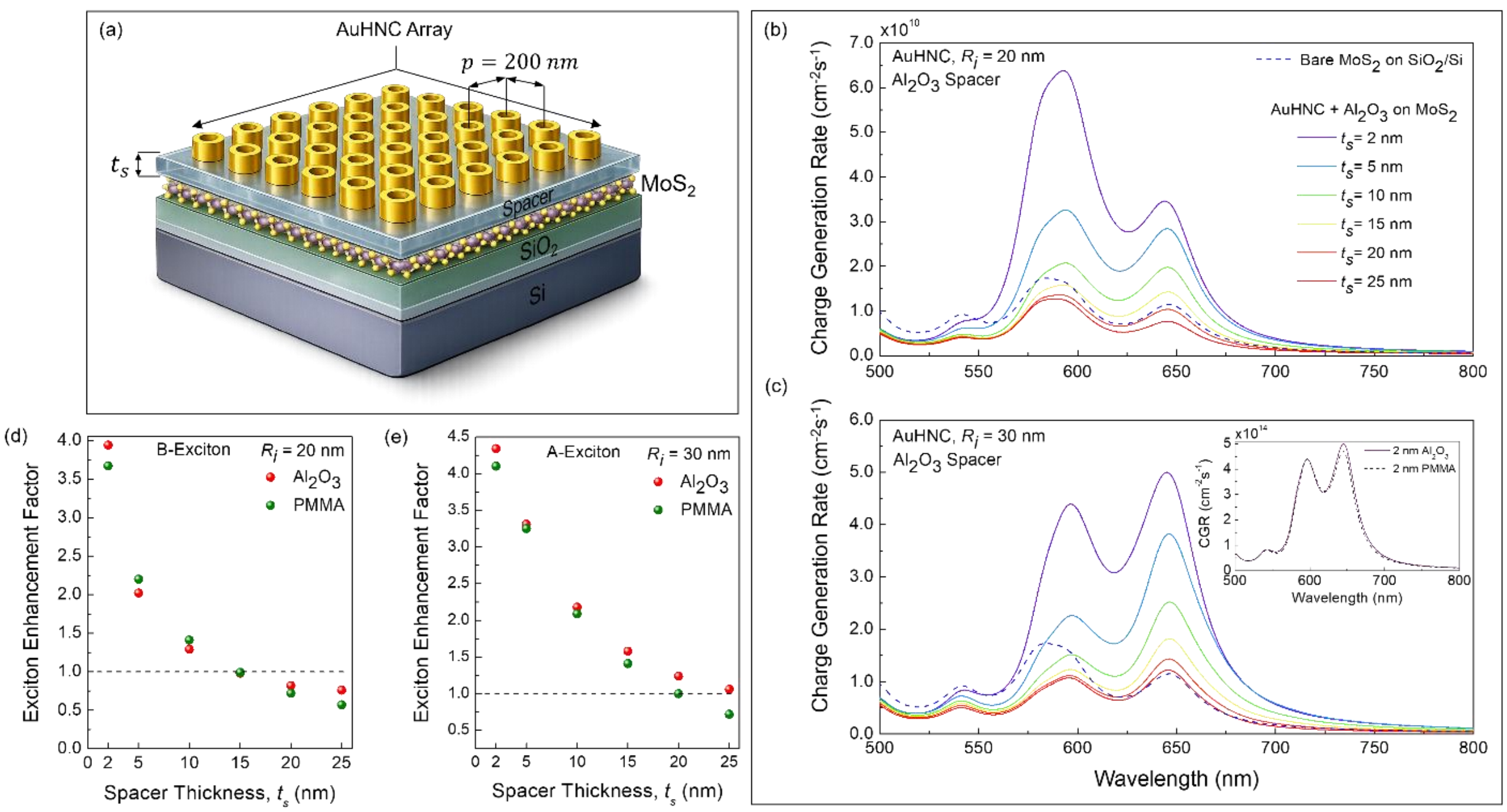

**Figure 3. Plasmon-enhanced excitonic charge generation in AuHNC-$MoS_2$ hybrid nanostructures. (a)** Schematic of the simulated periodic Au hollow nanocylinder (AuHNC) array with an in-plane periodicity of $p = 200$ nm, separated from monolayer $MoS_2$ by a dielectric spacer of thickness $t_s$ and supported on a $SiO_2$/Si substrate. **(b)** Charge generation rate (*CGR*) spectra of the AuHNC-$MoS_2$ hybrid system for the $R_i = 20$ nm geometry with $Al_2O_3$ spacer thicknesses ranging from $t_s = 2$ to 25 nm. The dashed curve corresponds to bare $MoS_2$ on the $SiO_2$/Si substrate. **(c)** *CGR* spectra of the AuHNC–$MoS_2$ hybrid system for the $R_i = 30$ nm geometry with $Al_2O_3$ spacer thicknesses ranging from $t_s = 2$ to 25 nm. The inset compares the CGR spectra obtained with 2 nm $Al_2O_3$ and PMMA spacers, highlighting the influence of the spacer dielectric environment on the plasmon-exciton spectral overlap. **(d)** B-exciton *CGR* enhancement factor extracted from the $R_i = 20$ nm geometry as a function of spacer thickness for $Al_2O_3$ and PMMA spacers. The enhancement reaches 3.94 for $Al_2O_3$ and 3.67 for PMMA at $t_s = 2$ nm, followed by a gradual decrease toward unity with increasing spacer thickness due to the spatial decay of the plasmonic near field. **(e)** A-exciton *CGR* enhancement factor extracted from the $R_i = 30$ nm geometry as a function of spacer thickness. At $t_s = 2$ nm, the enhancement reaches 4.34 for $Al_2O_3$ and 4.10 for PMMA, and progressively decreases for thicker spacers as the $MoS_2$ layer is displaced from the region of strongest near-field confinement.

We evaluate the resulting QE, PLE and exciton intensity redistribution of the AuHNC-$MoS_2$ hybrid system in **Figure 4**. **Figure 4a** demonstrates the QE maps of AuHNC-$MoS_2$ hybrid system with $R_i$ = 20 nm and $R_i$ = 30 nm and $Al_2O_3$ spacer (See **Supplementary Figure S4** for the PMMA counterpart). For the B-exciton targeted geometry ($R_i$ = 20, shown in green color map), the QE is maximized near the intrinsic excitonic transition wavelengths of ML $MoS_2$, reaching the maximum value of 0.54 at 612 nm for 25 nm spacer thickness indicating that the plasmonic resonance efficiently couples to the emission channel of the B-exciton. Similarly, red colormap of **Figure 4a** the QE map for A-exciton targeted geometry ($R_i$ = 30 nm and $Al_2O_3$ spacer). The maximum QE is recorded as 0.73 at 620 nm for 25 nm spacer thickness indicating stronger excitation enhancement compared with the B-exciton-targeted geometry in conjunction with the exciton enhancement factors reported in **Figure 3**. Note that the reported QE values represent the internal quantum efficiency of the hybrid system, not the external PL efficiency. As spacer thickness increases, non-radiative energy transfer to the metal is reduced, leading to higher QE values. However, the normalized radiative power simultaneously decreases with increasing separation and can undergo 6-fold reduction with respect to the bare ML $MoS_2$ radiative rate (**Table 2**). This highlights the need to optimize spacer thickness to balance radiative enhancement and non-radiative losses. The resulting PL enhancement factors are displayed in **Figure 4b** (and **Table 2**), which shows the calculated PL enhancement for the B-exciton and A-exciton transitions, respectively. The AuHNC-$MoS_2$ hybrid structures exhibit significantly larger enhancement compared to previously reported plasmonic systems such as Au nanorods (~3 times) and nanospheres (~45 times)[29,31-33]. This improvement originates from the strong electric-field confinement within the hollow nanocavity, which simultaneously enhances the excitation rate and the radiative decay channel. An additional advantage of employing AuHNCs with tunable aspect ratios is their ability to modify the relative spectral contributions of the A and B excitonic transitions.

To quantify this effect, we evaluate the normalized excitonic peak ratio (*NEPR*), defined as the ratio between the A and B excitonic peak intensities in the enhanced spectra normalized to the corresponding peak ratio of bare $MoS_2$. This metric therefore measures how the plasmonic cavity redistributes the excitonic emission balance relative to the intrinsic response of ML $MoS_2$. The resulting *NEPR*s are summarized in **Table 2** and plotted in **Figure 4c** for $Al_2O_3$ and PMMA spacers, respectively. For the $R_i$ = 30 nm geometry, where the plasmonic resonance is aligned closer to the A-exciton transition, the normalized ratio reaches values up to 2.4 for a 5 nm $Al_2O_3$ spacer, indicating that the A-exciton peak becomes more than twice as dominant relative to the B-exciton compared with the bare $MoS_2$ spectrum.

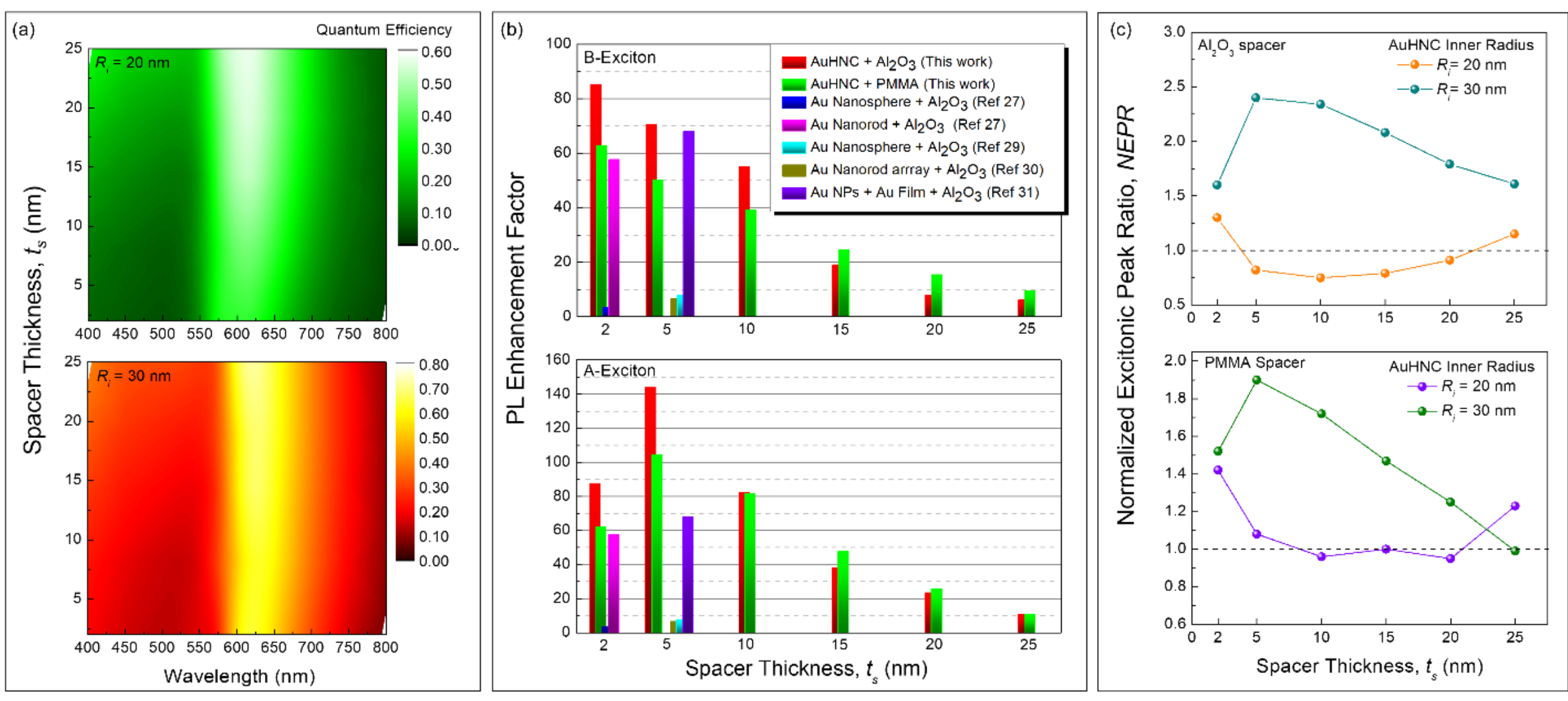

**Figure 4. Quantum efficiency, PL enhancement and normalized excitonic peak ratios in AuHNC-MoS2 hybrid structures. (a)** *QE* maps as a function of wavelength and spacer thickness $t_s$ for AuHNC geometries with inner radii $R_i = 20$ nm and $R_i = 30$ nm ($Al_2O_3$ spacer). The maximum *QE* occurs in the spectral region of the excitonic transitions, confirming that optimal radiative efficiency is achieved when the plasmonic resonance is aligned with the intrinsic A and B exciton energies **(b)** *PLE* factors at the B- and A-exciton wavelengths compared with previously reported plasmonic structures, shown as a function of spacer thickness for AuHNC structures with $Al_2O_3$ and PMMA spacers. Hollow nanocavities enable substantially larger *PLE* values reaching up to 145-fold in this work, representing up to 45-fold higher enhancement compared with previously reported plasmonic architectures. **(c)** *NEPR* plotted as a function of spacer thickness for $Al_2O_3$ (top) and PMMA (bottom) spacers for AuHNC inner radii $R_i = 20$nm and $R_i = 30$nm. Implemented cavity geometry enables controlled redistribution of the relative A and B excitonic contributions, showing that plasmonic nanocavities can spectrally reshape excitonic emission rather than merely amplify it.

A similar but slightly weaker redistribution is observed for the PMMA spacer where the ratio reaches a maximum of 1.9 across the investigated spacer thicknesses. In contrast, the $R_i = 20$ nm geometry, designed to interact more strongly with the B-exciton transition, yields normalized ratios closer to unity, indicating a more balanced enhancement of the two excitonic channels. These results show that AuHNC cavities do not simply provide broadband enhancement but instead reshape the relative excitonic emission spectrum through geometry-dependent plasmon-exciton alignment. Such control over the relative intensities of excitonic transitions provides a practical route toward spectral engineering of excitonic emission in atomically thin semiconductors.

**Table 2 | Geometry- and spacer-dependent excitonic enhancement metrics in AuHNC-$MoS_2$ hybrid structures.** Spacer-thickness-dependent excitation enhancement factors ($G/G_0$), normalized radiative decay rates ($F_{rad}$), PL enhancement (PLE), and the NEPR for the AuHNC–ML $MoS_2$ hybrid system. Results are reported for cavity geometries with inner radii $R_i = 30$nm (A-exciton-aligned) and $R_i = 20$nm (B-exciton-aligned) and for both $Al_2O_3$ and PMMA spacer layers. The NEPR metric quantifies the redistribution of the relative A/B excitonic peak intensities compared with bare ML-$MoS_2$.

| $R_i = 30\ nm$ (A-Exciton) | Excitation Enhancement Factor ($\frac{G}{G_0}$) | | Normalized Radiative Decay Rate ($F_{rad}$) | | PL Enhancement (*PLE*) | | Normalized excitonic peak ratio, (*NEPR*) | |
|---|---|---|---|---|---|---|---|---|
| Spacer Thickness (nm) | $Al_2O_3$ | PMMA | $Al_2O_3$ | PMMA | $Al_2O_3$ | PMMA | $Al_2O_3$ | PMMA |
| 2 | 4.34 | 4.1 | 19.12 | 15.12 | 82.98 | 61.99 | 1.6 | 1.52 |
| 5 | 3.31 | 3.25 | 43.46 | 32.03 | 143.85 | 104.10 | 2.4 | 1.9 |
| 10 | 2.18 | 2.09 | 37.72 | 39.08 | 82.23 | 81.68 | 2.34 | 1.72 |
| 15 | 1.58 | 1.41 | 23.97 | 33.75 | 37.87 | 47.59 | 2.08 | 1.47 |
| 20 | 1.24 | 1 | 18.75 | 25.84 | 23.25 | 25.84 | 1.79 | 1.25 |
| 25 | 1.06 | 0.72 | 10.03 | 14.61 | 10.63 | 10.52 | 1.61 | 0.99 |
| $R_i = 20\ nm$ (B-Exciton) | | | | | | | | |
| Spacer Thickness (nm) | $Al_2O_3$ | PMMA | $Al_2O_3$ | PMMA | $Al_2O_3$ | PMMA | $Al_2O_3$ | PMMA |
| 2 | 3.94 | 3.67 | 22.15 | 17.85 | 87.27 | 65.51 | 1.3 | 1.42 |
| 5 | 2.02 | 2.2 | 34.07 | 23.17 | 68.82 | 50.97 | 0.82 | 1.08 |
| 10 | 1.29 | 1.41 | 41.72 | 28.28 | 53.82 | 39.87 | 0.75 | 0.96 |
| 15 | 0.98 | 0.99 | 19.13 | 25.45 | 18.75 | 25.20 | 0.79 | 1 |
| 20 | 0.82 | 0.72 | 9.28 | 21.17 | 7.61 | 15.24 | 0.91 | 0.95 |
| 25 | 0.76 | 0.57 | 8.13 | 16.55 | 6.18 | 9.43 | 1.15 | 1.23 |

The orientation-averaged radiative enhancement was obtained by calculating $F_{\rm rad}$ for both $y$- and $z$-oriented (in-plane directions) dipoles and averaging their contributions. Although the present analysis focuses only on neutral excitonic transitions due to the band structure of the layer, the broad LSPR linewidth of the AuHNC suggests that trapped or localized excitonic states within the same spectral window could also couple efficiently to the plasmonic mode[34], consistent with the QY maps that reveal enhanced values over extended spectral regions and large areas near the nanocavity as shown in **Figure 4a**.

Overall, the combined control of cavity geometry, spacer thickness, and dielectric environment enables selective redistribution of excitonic emission and establishes the AuHNC- $MoS_2$ hybrid architecture as a tunable platform for exciton-specific light-matter interaction engineering.

## Conclusion, Discussion and Outlook

In this work, we demonstrate that vertically oriented AuHNCs provide an effective and tunable platform for enhancing excitonic emission in ML $MoS_2$ through controlled plasmon-exciton interactions. By engineering cavity geometry, the primary LSPR can be spectrally aligned with either the A or B excitonic transitions while maintaining operation in the weak-coupling regime. Systematic variation of the spacer thickness (2-25 nm) allows deterministic control of the plasmonic near-field overlap with the $MoS_2$ layer. For small separations, strong excitation enhancement is observed, with charge generation rates reaching 4.34 and 3.94 times the bare $MoS_2$/substrate values for the A and B excitonic transitions, respectively. As the spacer thickness increases, the near-field interaction decays rapidly and the excitation enhancement approaches or falls below the bare $MoS_2$ limit beyond approximately 20-25 nm.

Importantly, our results show that excitation enhancement alone does not directly translate into increased PL. Instead, the overall PL enhancement is determined by the balance between plasmon-enhanced excitation and cavity-modified radiative decay. Under optimized cavity and spacer conditions, the AuHNC-$MoS_2$ hybrid system achieves *PLE* of up to 143.85-fold for the A excitonic transition and 87.27-fold for the B-excitonic transition, exceeding enhancement levels typically reported for solid Au nanorods and nanospheres. In addition, the tunable cavity aspect ratio enables controlled redistribution of excitonic emission channels, producing *NEPR*s as high as 2.4 relative to bare $MoS_2$.

These results highlight the advantages of hollow coaxial plasmonic geometries, where hybrid radial-longitudinal modes provide strong field confinement and efficient local density of optical states engineering. Beyond fundamental plasmon-exciton coupling studies, this geometry-dependent design strategy offers a practical route for enhancing and spectrally shaping emission from low-quantum-yield 2D semiconductors. Such capabilities are particularly relevant for the development of ultrathin light sources, on-chip nanophotonic emitters, valleytronic devices, and integrated optoelectronic platforms based on atomically thin materials[35-38].

## Methodology

To accurately capture both the atomistic excitonic response of ML $MoS_2$ and the electromagnetic field confinement of the plasmonic cavity, we combined first-principles excitonic calculations with full-wave electromagnetic simulations. The optical response of the $MoS_2$-Au hollow nanocavity system was simulated using the FDTD method.

The optical response of the $MoS_2$-spacer-AuHNC system was simulated using the finite difference time-domain (FDTD) method, which numerically solves Maxwell's equations in both space and time. In this approach, Maxwell's curl equations

$$\nabla \times \mathbf{E} = -\mu \frac{\partial \mathbf{H}}{\partial t} \tag{5}$$

$$\nabla \times \mathbf{H} = \varepsilon \frac{\partial \mathbf{E}}{\partial t} + \mathbf{J} \tag{6}$$

are discretized on a Yee grid, where **E** and **H** denote the electric and magnetic fields, μ is the magnetic permeability, ε is the permittivity, and **J** is the induced current density. Perfectly matched layers (PMLs) were applied at the simulation boundaries to emulate an open electromagnetic environment and suppress artificial reflections. All simulations were performed using a three-dimensional FDTD solver (Ansys Lumerical FDTD) with spatial and temporal discretization chosen to satisfy the Courant stability condition while resolving the nanoscale features of the $MoS_2$ layer and the plasmonic cavity.

The dispersive optical properties of the materials were implemented using experimentally validated refractive-index datasets. Gold was described using the Johnson and Christy data set[39], while Si, $SiO_2$, and $Al_2O_3$ were modeled using wavelength-dependent optical constants from the Palik Handbook[40-42]. PMMA was implemented as a transparent dielectric using its Sellmeier-type dispersion relation [43]:

$$n^2(\lambda) = 1 + \frac{0.99654\lambda^2}{\lambda^2 - 0.00787} + \frac{0.18964\lambda^2}{\lambda^2 - 0.02191} + \frac{0.00411\lambda^2}{\lambda^2 - 3.85727} \tag{7}$$

which accurately reproduces its refractive index across the visible spectrum.

To evaluate the optical response of the plasmonic cavity, broadband extinction spectra were calculated using a total-field scattered-field (TFSF) plane-wave source spanning 400-800 nm. Absorption within the Au nanocavity and spacer layers was obtained by integrating the power dissipation density over their volumes, while scattering was calculated from the far-field power exiting the TFSF boundaries.

The optical response of ML $MoS_2$ was obtained from first-principles calculations using the density functional theory (DFT)-Bethe-Salpeter equation (BSE) framework. Ground-state electronic structure calculations were performed within the projector augmented-wave (PAW) method using the generalized gradient approximation (GGA) in the PBE form with a plane-wave cutoff of 500 eV and a 16 × 16 × 1 Γ-centered k-point mesh[44, 45]. A vacuum spacing of 20 Å was applied along the out-of-plane direction to avoid interlayer interactions. Spin-orbit coupling was included to correctly capture the splitting between the A and B excitons. Quasiparticle corrections were calculated using the $G_0W_0$ approximation, followed by excitonic calculations within the BSE framework[46-48]. An energy cutoff of 150 eV was used for the response function.

The resulting complex dielectric function

$$\varepsilon(\omega) = \varepsilon_1(\omega) + i\varepsilon_2(\omega) \tag{8}$$

was implemented in the FDTD simulations as a surface conductivity layer according to

$$\sigma(\omega) = -i\omega t\varepsilon_0[\varepsilon(\omega) - 1] \tag{9}$$

where $\varepsilon_0$ is the vacuum permittivity, $\omega$ is the angular frequency, and $t$is the effective ML thickness. The $MoS_2$ layer was modeled as an in-plane isotropic sheet conductivity, consistent with the dominant in-plane excitonic response of ML TMDCs.

*CGR*s were calculated from the spatially resolved electromagnetic fields. Under optical excitation, the local absorption in the $MoS_2$ layer is given by the power dissipation density

$$P(\mathbf{r}, \lambda) = \frac{1}{2}\omega\varepsilon_0\varepsilon_2(\mathbf{r}, \lambda) \mid \mathbf{E}(\mathbf{r}, \lambda) \mid^2 \tag{10}$$

The corresponding electron–hole pair generation rate is

$$G(\mathbf{r}, \lambda) = \frac{P(\mathbf{r},\lambda)}{\hbar\omega} = \frac{1}{2}\frac{\omega\varepsilon_0\varepsilon_2|\mathbf{E}|^2}{\hbar\omega} \tag{11}$$

The generation maps were integrated over the $MoS_2$ plane to obtain wavelength-dependent CGR spectra.

Radiative and non-radiative decay rates were evaluated by modeling the $MoS_2$ exciton as an in-plane oscillating electric dipole located near the plasmonic cavity. In free space, the intrinsic dipole decay rate is given by [49]:

$$\gamma_0 = \frac{\omega^3|p|^2}{3\pi\varepsilon_0\hbar c^3} \tag{12}$$

where $p$ is the dipole moment and $c$ is the speed of light. Within the cavity environment, the total decay rate was obtained from the total emitted power

$$\gamma_{tot} = \frac{P_{tot}}{P_0}\gamma_0 \tag{13}$$

where $P_0$ is the dipole power in free space. Radiative emission was determined using far-field monitors,

$$\gamma_{rad} = \frac{P_{rad}}{P_0}\gamma_0 \tag{14}$$

while the non-radiative decay rate associated with metallic losses was obtained as:

$$\gamma_{nonrad} = \gamma_{tot} - \gamma_{rad}. \tag{15}$$

Near-field monitors were used to capture the full Poynting-vector flux surrounding the emitter,

$$P_{tot} = \oint \mathbf{S} \cdot dA \tag{16}$$

whereas far-field projection monitors were used to evaluate

$$P_{rad} = \oint_{far} \mathbf{S} \cdot dA. \tag{17}$$

This framework allows a clear separation of radiative enhancement and non-radiative plasmonic losses and enables quantitative evaluation of cavity-modified decay dynamics.

**Acknowledgments**

E.O.P. acknowledges financial support from the Scientific and Technological Research Council of Türkiye (TÜBİTAK) through Grant No. 222N308 within the CHIST-ERA program. Additional support was provided by the Turkish Academy of Sciences Outstanding Young Scientists Awards (GEBİP) 2025.

**Author Contributions**

A.E.Y. performed numerical simulations, carried out the calculations, and extracted the raw data underlying the reported results. A.E.Y. also contributed to the preparation of the manuscript. E.O.P. conceived the research idea, supervised the project, and led the preparation of the manuscript, including the development of the final figures and the main text. All authors discussed the results and contributed to the final version of the manuscript.

**Competing Interests**

The authors declare no competing financial or non-financial interests.

**Data Availability**

The data required to evaluate the results in this article are tabulated and included in the main article and its Supplementary Information. All the simulation settings and analysis procedures are described and given in detail in the Methodology section.

**References**

1. Wang, G. *et al. Colloquium*: Excitons in atomically thin transition metal dichalcogenides. *Reviews of Modern Physics* vol. 90 (2018).
2. He, K. *et al.* Tightly Bound Excitons in Monolayer WSe2. *Physical Review Letters* vol. 113 (2014).
3. Mak, K., Lee, C., Hone, J., Shan, J. & Heinz, T. Atomically Thin MoS2: A New Direct-Gap Semiconductor. *Physical Review Letters* vol. 105 (2010).
4. Splendiani, A. *et al.* Emerging Photoluminescence in Monolayer MoS2. *Nano Letters* vol. 10 1271–1275 (2010).
5. Xiao, D., Liu, G., Feng, W., Xu, X. & Yao, W. Coupled Spin and Valley Physics in Monolayers of MoS2 and Other Group-VI Dichalcogenides. *Physical Review Letters* vol. 108 (2012).
6. Lopez-Sanchez, O., Lembke, D., Kayci, M., Radenovic, A. & Kis, A. Ultrasensitive photodetectors based on monolayer MoS2. *Nature Nanotechnology* vol. 8 497–501 (2013).
7. Ross, J. *et al.* Electrical control of neutral and charged excitons in a monolayer semiconductor. *Nature Communications* vol. 4 (2013).
8. Zeng, H., Dai, J., Yao, W., Xiao, D. & Cui, X. Valley polarization in MoS2 monolayers by optical pumping. *Nature Nanotechnology* vol. 7 490–493 (2012).
9. Zheng, W. *et al.* Light Emission Properties of 2D Transition Metal Dichalcogenides: Fundamentals and Applications. *Advanced Optical Materials* vol. 6 (2018).
10. Ansari, N. & Ghorbani, F. Light absorption optimization in two-dimensional transition metal dichalcogenide van der Waals heterostructures. *Journal of the Optical Society of America B-Optical Physics* vol. 35 1179–1185 (2018).
11. Amani, M. *et al.* Near-unity photoluminescence quantum yield in MoS2. *Science* vol. 350 1065–1068 (2015).
12. Eda, G. & Maier, S. Two-Dimensional Crystals: Managing Light for Optoelectronics. *ACS Nano* vol. 7 5660–5665 (2013).

13. Sundaram, R. *et al.* Electroluminescence in Single Layer MoS2. *Nano Letters* vol. 13 1416–1421 (2013).
14. Lee, B. *et al.* Fano Resonance and Spectrally Modified Photoluminescence Enhancement in Monolayer MoS2 Integrated with Plasmonic Nanoantenna Array. *Nano Letters* vol. 15 3646–3653 (2015).
15. Lunnemann, P. & Koenderink, A. The local density of optical states of a metasurface. *Scientific Reports* vol. 6 (2016).
16. Jin, S. *et al.* Plasmonic tuning of dark-exciton radiation dynamics and far-field emission directionality in monolayer WSe2. *Science Advances* vol. 12 (2026).
17. Bharadwaj, P., Deutsch, B. & Novotny, L. Optical Antennas. *Advances in Optics and Photonics* vol. 1 438–483 (2009).
18. Tame, M. *et al.* Quantum plasmonics. *Nature Physics* vol. 9 329–340 (2013).
19. Butun, S., Tongay, S. & Aydin, K. Enhanced Light Emission from Large-Area Monolayer MoS2 Using Plasmonic Nanodisc Arrays. *Nano Letters* vol. 15 2700–2704 (2015).
20. Moreau, A. *et al.* Controlled-reflectance surfaces with film-coupled colloidal nanoantennas. *Nature* vol. 492 86-+ (2012).
21. Zhang, W., Yang, J. & Lu, X. Tailoring Galvanic Replacement Reaction for the Preparation of Pt/Ag Bimetallic Hollow Nanostructures with Controlled Number of Voids. *ACS Nano* vol. 6 7397–7405 (2012).
22. Sun, Y. & Xia, Y. Shape-controlled synthesis of gold and silver nanoparticles. *Science* vol. 298 2176–2179 (2002).
23. Thickness-Dependent Refractive Index of 1L, 2L, and 3L MoS2, MoSe2, WS2, and WSe2 - Hsu Advanced Optical Materials - Wiley Online Library (2019).
24. Stewart, M. *et al.* Multispectral Thin Film Biosensing and Quantitative Imaging Using 3D Plasmonic Crystals. *Analytical Chemistry* vol. 81 5980–5989 (2009).
25. Kaminska, I. *et al.* Near-Field and Far-Field Sensitivities of LSPR Sensors. *Journal of Physical Chemistry C* vol. 119 9470–9476 (2015).
26. Baumberg, J. J., Aizpurua, J., Mikkelsen, M. H. & Smith, D. R. Extreme nanophotonics from ultrathin metallic gaps. *Nat. Mater.* **18**, 668–678 (2019).
27. Hou, S. *et al.* Manipulating Coherent Light–Matter Interaction: Continuous Transition between Strong Coupling and Weak Coupling in $MoS_2$ Monolayer Coupled with Plasmonic Nanocavities. *Adv. Opt. Mater.* **7**, 1900857 (2019).
28. Wei, H. *et al.* Plasmon–Exciton Interactions: Spontaneous Emission and Strong Coupling. *Adv. Funct. Mater.* **31**, 2100889 (2021).
29. Garai, M., Zhu, Z., Shi, J., Li, S. & Xu, Q. Single-particle studies on plasmon enhanced photoluminescence of monolayer MoS2 by gold nanoparticles of different shapes. *Journal of Chemical Physics* vol. 155 (2021).
30. Zhao, W. *et al.* Exciton-Plasmon Coupling and Electromagnetically Induced Transparency in Monolayer Semiconductors Hybridized with Ag Nanoparticles. *Advanced Materials* vol. 28 2709–2715 (2016).

31. Qi, X. *et al.* Effects of gap thickness and emitter location on the photoluminescence enhancement of monolayer MoS2 in a plasmonic nanoparticle-film coupled system. *Nanophotonics* vol. 9 2097–2105 (2020).
32. Lee, K. *et al.* Plasmonic Gold Nanorods Coverage Influence on Enhancement of the Photoluminescence of Two-Dimensional MoS2 Monolayer. *Scientific Reports* vol. 5 (2015).
33. Huang, Y. *et al.* High-enhancement photoluminescence of monolayer MoS 2 in hybrid plasmonic systems. *Applied Optics* vol. 63 2704–2709 (2024).
34. Wang, Q. *et al.* Resonance Tuning of Localized Excitons via a Plasmonic Nanocavity. *ACS Nano* **20**, 6279–6286 (2026).
35. Nielsen, K. *et al.* Programmable nonlinear quantum photonic circuits. *Nature Communications* vol. 16 (2025).
36. Polat, E. *et al.* Graphene-Enabled Optoelectronics on Paper. *ACS Photonics* vol. 3 964–971 (2016).
37. Polat, E. *et al.* Flexible graphene photodetectors for wearable fitness monitoring. *Science Advances* vol. 5 (2019).
38. Tyulnev, J. et al, I. ,. Jiménez-Galán, Á. ,. Poborska. Valleytronics in bulk MoS2 with a topologic optical field. *Nature* vol. 628 690–690 (2024).
39. Johnson, P. B. & Christy, R. W. Optical Constants of the Noble Metals. *Physical Review B* vol. 6 4370–4379 (1972).
40. Edwards, D. F. Silicon (Si). *Handbook of Optical Constants of Solids* 547–569 (1985)
41. Gervais, F. Aluminum Oxide (Al2O3). *Handbook of Optical Constants of Solids* 761–775 (1998)
42. Philipp, H. R. Silicon Dioxide (SiO2), Type α (Crystalline). *Handbook of Optical Constants of Solids* 719–747 (1985)
43. Sultanova, N., Kasarova, S. & Nikolov, I. Dispersion Properties of Optical Polymers. *Acta Physica Polonica a* vol. 116 585–587 (2009).
44. Blochl, P. Projector Augmented-Wave Method *Physical Review B* vol. 50 17953–17979 (1994).
45. Perdew, J., Burke, K. & Ernzerhof, M. Generalized gradient approximation made simple. *Physical Review Letters* vol. 77 3865–3868 (1996).
46. Hybertsen, M. S. & Louie, S. G. Electron correlation in semiconductors and insulators: Band gaps and quasiparticle energies. *Physical Review B* vol. 34 5390–5413 (1986).
47. Shishkin, M. & Kresse, G. Self-consistent *GW* calculations for semiconductors and insulators. *Physical Review B* vol. 75 (2007).
48. Hedin, L. New Method for Calculating the One-Particle Green's Function with Application to the Electron-Gas Problem. *Physical Review* vol. 139 A796–A823 (1965).
49. Gulucu, A. & Polat, E. Optically Switchable Fluorescence Enhancement at Critical Interparticle Distances. *Advanced Theory and Simulations* vol. 8 (2025).

**Supplementary Information**

**for**

# Geometry-Controlled Exciton Selectivity in Monolayer $MoS_2$ Using Plasmonic Hollow Nanocavities

Abdullah Efe Yildiz[1,2] and Emre Ozan Polat[1,2,*]

[1]*Department of Physics, Bilkent University, 06800, Ankara, Turkey*

[2]*UNAM - National Nanotechnology Research Center and Institute of Materials Science and Nanotechnology, Bilkent University, Ankara 06800, Turkey*

**Corresponding author: emre.polat@bilkent.edu.tr*

The supplementary figures provide additional numerical results supporting the plasmon-exciton interaction analysis presented in the main text.

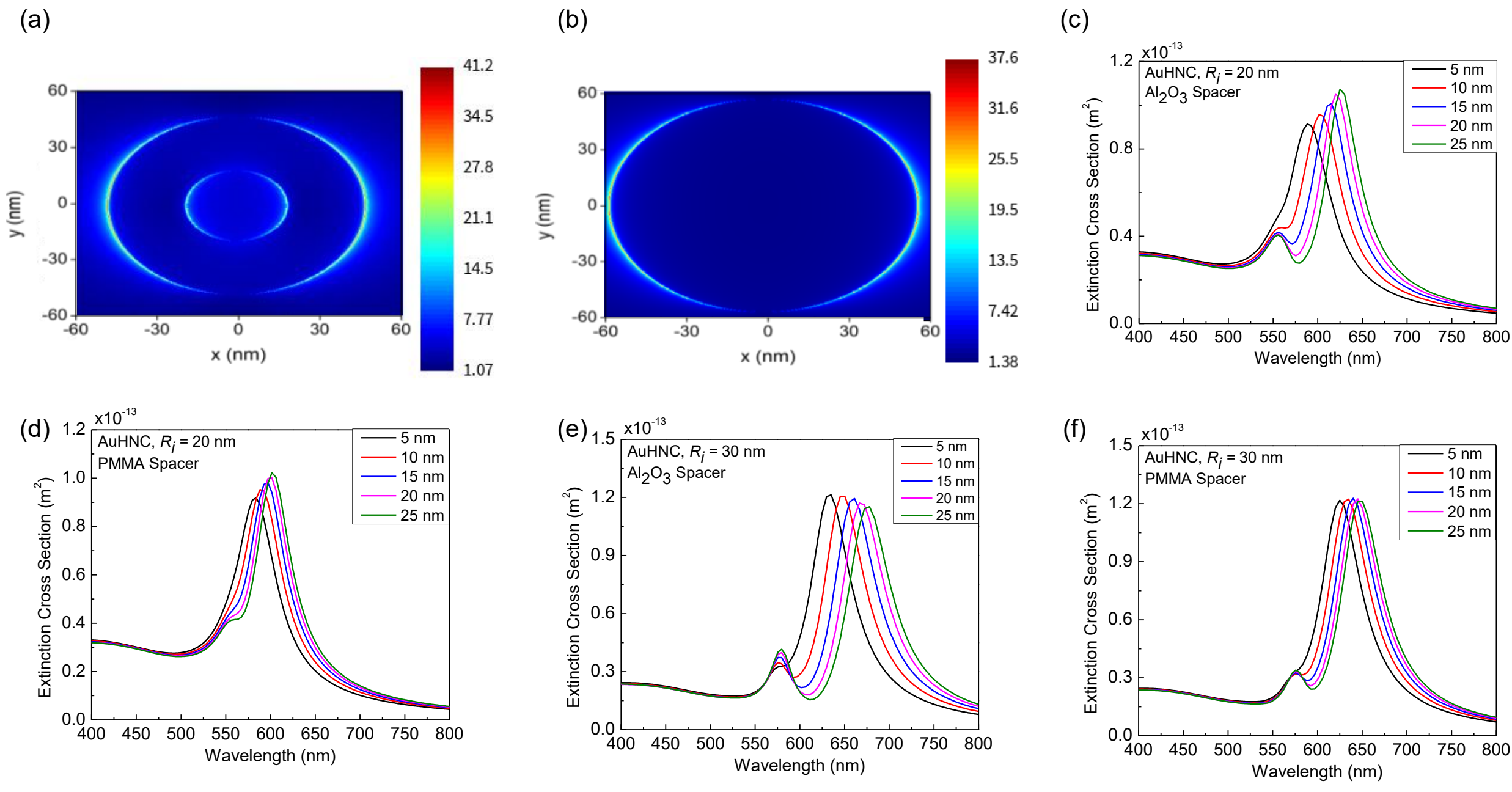

**Supplementary Figure S1. Electric-field confinement and geometry-dependent plasmonic response of hollow nanocavities. (a)** Simulated local electric-field intensity enhancement $(|E|/|E_0|)^2$ in the x-y plane for a hollow gold nanocylinder (AuHNC), showing strong field localization at both the inner and outer cavity boundaries arising from hybridized radial-longitudinal plasmon modes. **(b)** Corresponding electric-field distribution for a solid gold nanocylinder, where the field is primarily confined along the outer perimeter, illustrating the reduced modal confinement compared with the hollow structure. **(c–f)** Extinction spectra for selected geometries corresponding to different cavity aspect ratios (*CAR*s) designed to spectrally align the plasmon resonance with the B **(c,d)** and A **(e,f)** excitonic transitions of monolayer $MoS_2$ for $Al_2O_3$ and PMMA spacer environments.

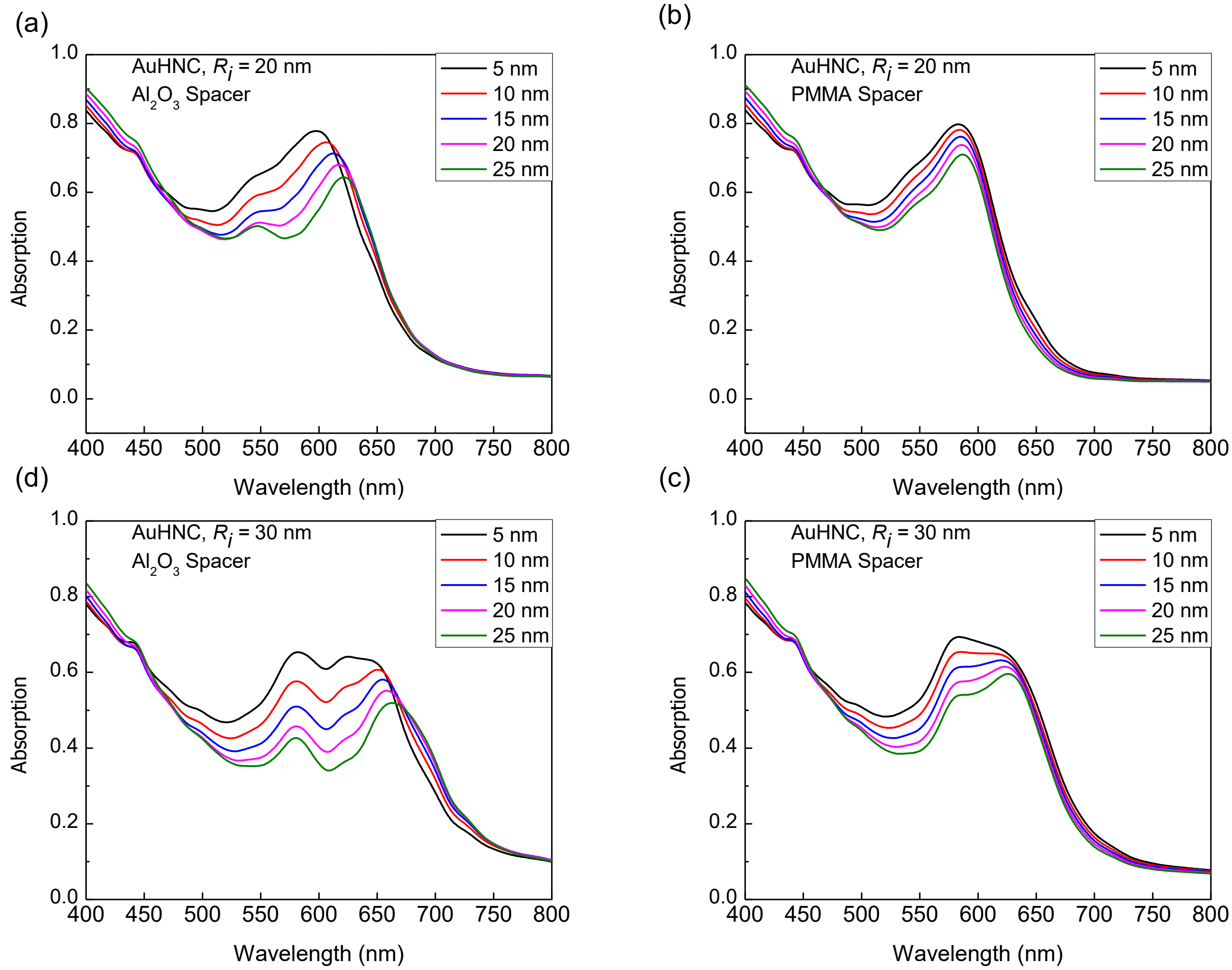

**Supplementary Figure S2. Spacer-dependent absorption response of AuHNC-$MoS_2$ hybrid structures. (a)** Absorption spectra of the hybrid structure with an $Al_2O_3$ spacer for geometries engineered to spectrally overlap with the B exciton of monolayer $MoS_2$. **(b)** Corresponding absorption spectra for a PMMA spacer targeting the B-exciton transition. **(c)** Absorption spectra for the $Al_2O_3$ spacer case with cavity geometries designed to align the plasmon resonance with the A exciton. **(d)** Corresponding absorption spectra for the PMMA spacer configuration targeting the A-exciton transition.

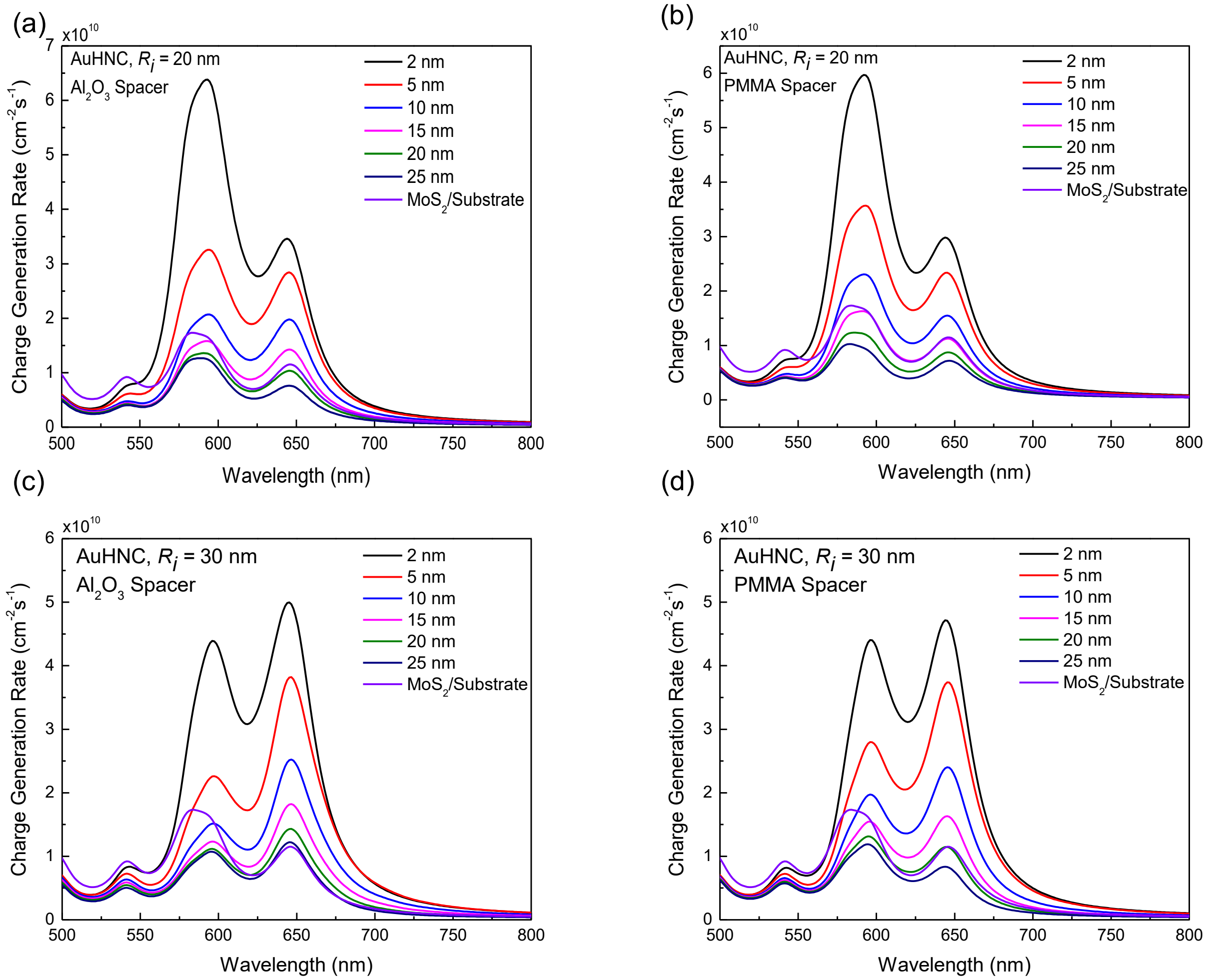

**Supplementary Figure S3. Charge-generation-rate (CGR) spectra of AuHNC-$MoS_2$ hybrid systems. (a)** CGR spectra for hybrid structures with an $Al_2O_3$ spacer designed to spectrally target the B exciton of monolayer $MoS_2$. **(b)** Corresponding CGR spectra for a PMMA spacer tuned to the B-exciton transition. **(c)** CGR spectra for the $Al_2O_3$ spacer configuration engineered to align the plasmonic resonance with the A exciton. **(d)** CGR spectra for the PMMA spacer configuration targeting the A-exciton transition. In each panel, the spacer thickness varies from 2 to 25 nm, demonstrating how the separation between the Au hollow nanocylinder and the $MoS_2$ layer modifies the near-field interaction and consequently the exciton charge-generation rate. The response of the bare $MoS_2$/substrate system is shown for comparison.

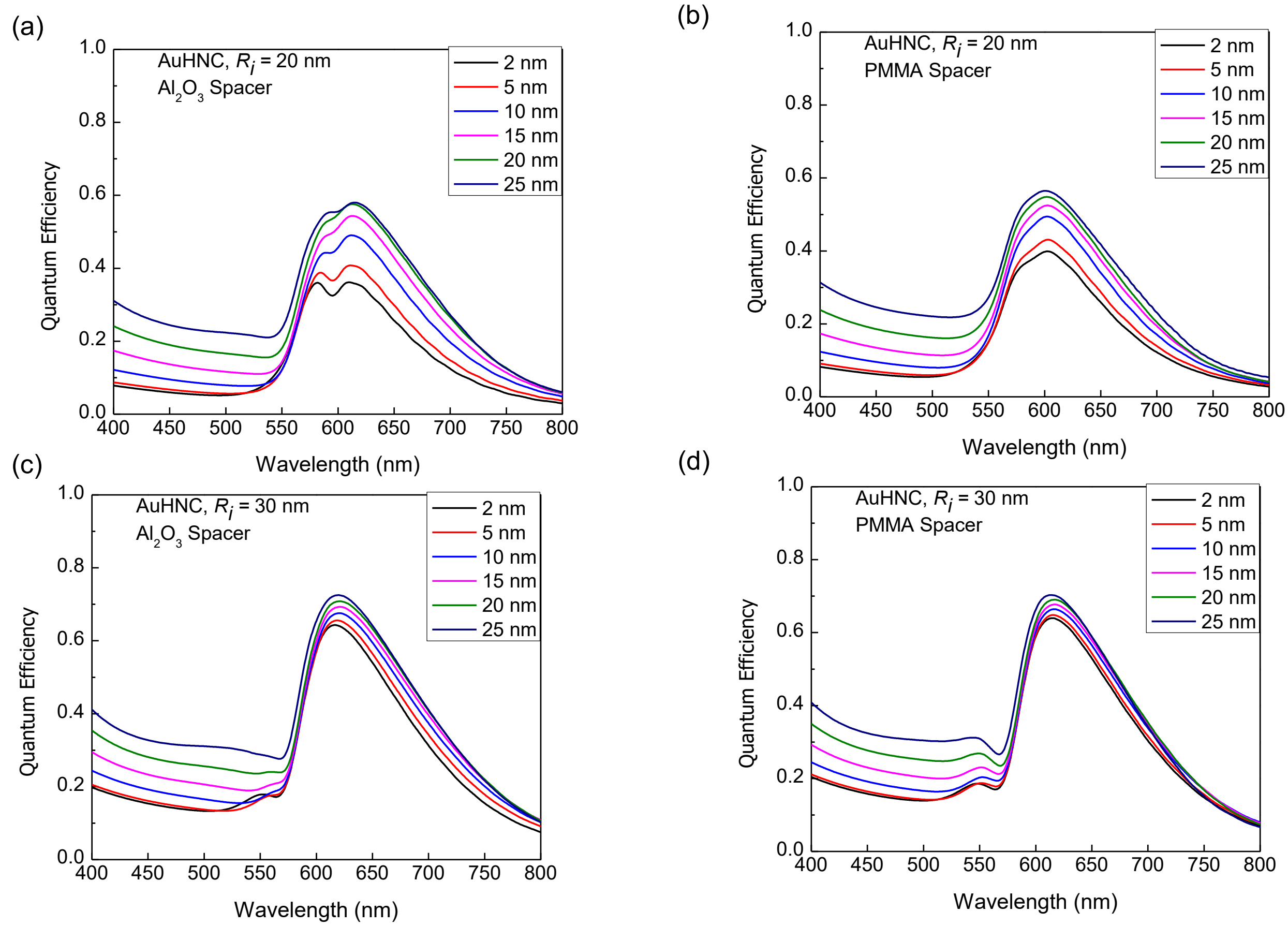

**Supplementary Figure S4. Spacer-dependent quantum efficiency of AuHNC-$MoS_2$ hybrid structures**. **(a,b)** Quantum-efficiency spectra for geometries designed to target the B exciton of monolayer $MoS_2$ with $Al_2O_3$ and PMMA spacers, respectively. **(c,d)** Corresponding quantum-efficiency spectra for geometries engineered to align the plasmonic resonance with the A exciton for $Al_2O_3$ and PMMA spacer layers. In each panel, the spacer thickness varies from 2 to 25 nm, illustrating how the separation between the hollow gold nanocylinder and the $MoS_2$ layer regulates the balance between radiative enhancement and non-radiative energy transfer to the metal.